\documentclass[apj]{emulateapj}
\usepackage{graphicx,pdfpages,amsmath}
\usepackage{sidecap}
\usepackage{natbib}

\newcommand{\bjdtdb}{\ensuremath{\rm {BJD_{TDB}}}}
\newcommand{\feh}{\ensuremath{\left[{\rm Fe}/{\rm H}\right]}}
\newcommand{\teff}{\ensuremath{T_{\rm eff}}}

\newcommand{\msun}{\ensuremath{\,M_\Sun}}
\newcommand{\rsun}{\ensuremath{\,R_\Sun}}
\newcommand{\lsun}{\ensuremath{\,L_\Sun}}
\newcommand{\mj}{\ensuremath{\,M_{\rm J}}}
\newcommand{\rj}{\ensuremath{\,R_{\rm J}}}
\newcommand{\fave}{\langle F \rangle}
\newcommand{\fluxcgs}{10$^9$ erg s$^{-1}$ cm$^{-2}$}
\shorttitle{The DEdicated MONitor of EXotransits (DEMONEX): Seven Transits of XO-4b}
\shortauthors{S. Villanueva Jr. et al.}

\begin{document}

\title{The DEdicated MONitor of EXotransits (DEMONEX): Seven Transits of XO-4b}

\author{S. Villanueva Jr.}
\affil{Department of Astronomy, The Ohio State University, 140 West 18th Av., Columbus OH 43210}

\author{J. D. Eastman}
\affil{Harvard-Smithsonian Center for Astrophysics, 60 Garden St., Cambridge, MA 02138}

\and

\author{B. S. Gaudi}
\affil{Department of Astronomy, The Ohio State University, 140 West 18th Av., Columbus OH 43210}

\email{svillan@astronomy.ohio-state.edu}

\begin{abstract}
The DEdicated MONitor of EXotransits (DEMONEX) was a 20 inch robotic and automated telescope to monitor bright stars hosting transiting exoplanets to discover new planets and improve constraints on the properties of known transiting planetary systems. We present results for the misaligned hot Jupiter XO-4b containing 7 new transits from the DEMONEX telescope, including 3 full and 4 partial transits. We combine these data with archival light curves and archival radial velocity measurements to derive the host star mass $M_{*}=1.293_{-0.029}^{+0.030}~\msun$ and radius $R_{*}=1.554_{-0.030}^{+0.042}~\rsun$ as well as the planet mass $M_{P}=1.615_{-0.099}^{+0.10}~\mj$ and radius $R_{P}=1.317_{-0.029}^{+0.040}~\rj$ and a refined ephemeris of $P=4.1250687\pm0.0000024$~days and $T_{0}=2454758.18978\pm0.00024~\bjdtdb$. We include archival Rossiter-McLaughlin measurements of XO-4 to infer the stellar spin-planetary orbit alignment $\lambda=-40.0_{-7.5}^{+8.8}$~degrees.

We test the effects of including various detrend parameters, theoretical and empirical mass-radius relations, and Rossiter-McLaughlin models. We infer that detrending against CCD position and time or airmass can improve data quality, but can have significant effects on the inferred values of many parameters --- most significantly $R_{P}/R_{*}$ and the observed central transit times $T_{C}$. In the case of $R_{P}/R_{*}$ we find that the systematic uncertainty due to detrending can be three times that of the quoted statistical uncertainties. The choice of mass-radius relation has little effect on our inferred values of the system parameters. The choice of Rossiter-McLaughlin models can have significant effects of the inferred values of $v\sin{I_{*}}$ and the stellar spin-planet orbit angle $\lambda$.
\end{abstract}


\section{Introduction}

Hot Jupiters are an unique class of extrasolar planets (exoplanets) known for their proximity to their host stars and their large masses. With the discovery of the first hot Jupiter \citep{mayor1995}, we now know that understanding the physical nature of hot Jupiters can play a large role in constraining our understanding of a variety of theories regarding planetary formation, disk formation, and planet migration \citep{marcy2005,fabrycky2007,nagasawa2008}.

Hot Jupiters provide unique observational opportunities relative to exoplanets of other populations. The small semi-major axis of the orbits of hot Jupiters increase the a priori transit probability, this coupled with a large observational transit signal due to hot Jupiters' large relative radii, and their short periods provide many observational opportunities upon which to identify hot Jupiters and perform follow-up observations. These observational advantages are not unique to photometric observations, but also to radial velocity (RV) measurements where the large masses of hot Jupiters increase the semi-amplitude of the radial velocity signal as well as to Rossiter-McLaughlin (RM) measurements where the RM signal is proportional to the transit depth.

In general, the transits provide a number of geometric ratios to relate the star and planet (i.e., the depth of the transit is related to the planet/star radius ratio), as well as the density of the star. When combined with another constraint on the mass and radius of the primary (e.g., derived from isochrones using the spectroscopic stellar effective temperature and metallicity), and a measurement of the Doppler amplitude from radial velocity observations, it becomes possible to estimate the masses and radii of both the planet and star. Additional measurements allow for the derivation of a large number of physical parameters, not just the planet and stellar masses and radii \citep[see][]{winn2010}.

Following the discovery of systems containing hot Jupiters, one generally would like to follow-up with observations to better understand the properties of the host star and planet. In general this requires additional photometric and radial velocity observations. Providing stricter constraints on known hot Jupiters provides predictions for future observations, increasing the efficiency of those observations.

The purpose of this paper is three-fold in nature. First, to present new data on XO-4b from the DEdicated MONitor of EXotransits (DEMONEX) telescope. Second, we investigate the effects of choices of detrending variables, mass-radius relations, Rossiter-McLaughlin models, and other models or user defined variables on the derived parameters of the system. Third, we improve the parameters of XO-4 and XO-4b by globally fitting all available data in a homogeneous manner.

\section{Data}
The following sections contain summaries of all available data used in our global analysis as well as information regarding the data reduction process used for the new DEMONEX data.

\subsection{DEMONEX Observations of XO-4}

New observations of XO-4b were made using the DEdicated MONitor of EXotransits (DEMONEX) \citep{east2010}. DEMONEX was a low-cost, 0.5 meter, robotic telescope constructed from commercially-available parts operated out of Winer Observatory in Sonoita, Arizona. DEMONEX monitored bright stars hosting known transiting planets over a three year period from 2008-2011 in order to provide a homogeneous data set of precise relative photometry for over 40 transiting systems.

There are 20 nights of data from November 2008 to May 2010 taken during primary transits of XO-4b. All observations were made in the Sloan $z$ band. Due to issues with the mount, 5 nights were lost due to the mount pointing to the incorrect field. In addition, the DEMONEX observing strategy prioritized full transits of other targets over partial transits of XO-4b. Often, that resulted in small observing windows, which we opted to fill with observations near transit in case they happened to be useful rather than sitting idle. Unfortunately, 4 nights only got data out-of-transit and 4 nights only captured the flat bottom of the transit, neither of which ended up providing useful constraints. Thus we were left with an unfortunately low yield of 7/20 (35\%) usable nights, but we obtain 3 full transits, 2 ingresses, 2 egresses.

\subsection{McCullough 2008}

\citet{MC08} (MC08) report the original discovery of the planet XO-4b detected using the XO telescope \citep{MC06}. Follow-up $BVR$ observations were made by the XO Extended Team (XOET) as well as follow-up $R$ band photometric observations using the Perkins 1.8-m telescope. Follow-up spectroscopic measurements were made using the Harlan J. Smith 2.7-m telescope and the 11-m Hobby-Eberly Telescope. MC08 perform an analysis of the spectra to report the stellar properties of the host star including stellar effective temperature $T_{\rm{eff}}$, surface gravity $\log{g_*}$, metallicity \feh, projected rotational velocity $v\sin{I_{*}}$, and the RV semi-amplitude $K$. Combining the stellar parameters with the light curves MC08 report a planet mass of $M_{P}=1.72\pm0.20~M_{J}$, radius of $R_{P}=1.34\pm0.048~R_{J}$, orbital period of $P=4.12502\pm0.00002$~days, and heliocentric Julian date at mid-transit of $2454485.9322 \pm 0.0004$ for XO-4b.

We adopt the stellar properties, $T_{\rm{eff}}$, $\log{g_*}$, \feh, and $v\sin{I_{*}}$, from MC08 as Gaussian priors for our global analysis as described in \S\ref{datasec}. Additionally, the authors of MC08 kindly provided the original XO light curves, the $BVR$ band XOET follow-up light curves, and the Perkins follow-up light curves, which we use, along with the radial velocity data listed in the MC08 paper in our global analysis.

We convert the times from the MC08 XOET light curves from $\mathrm{HJD_{UT}}$ to \bjdtdb~using the IDL code {\tt HJD2BJD}\footnote{\emph{http://astroutils.astronomy.ohio-state.edu/time/hjd2bjd.html}} to maintain uniform time stamps across all available light curves.

\subsection{Narita 2010}\label{naritasec}

\citet{narita2010} (N10) report new Sloan $z$ band photometric and radial velocity observations of XO-4 conducted with the FLWO 1.2 m telescope (photometric) and the 8.2 m Subaru Telescope (RV). Based on these new light curves, N10 report a refined transit ephemeris for XO-4b of $P=4.1250828\pm0.0000040$~days and $T_{c}=2454485.93323\pm0.00039~\bjdtdb$.

N10 also report the first measurements of the Rossiter-McLaughlin effect of XO-4b. N10 estimate the sky-projected angle between the stellar spin axis an the planetary orbital axis to be $\lambda=-46.7_{-6.1}^{+8.1}$~degrees. We compare the N10 results from the publicly-available N10 light curves and radial velocity data against the results we obtain using the same data set outlined in \S\ref{compsec} to validate our light curve analysis methods and procedures. We also include the publicly-available N10 light curves and radial velocity data in our global analysis.

\subsection{Todorov 2012}

\citet{todorov2012} use the Spitzer Space Telescope \citep{spitzer} and Infrared Array Camera \citep{irac} to obtain $3.6~\mu m$ and $4.5~\mu m$ observations of the secondary eclipses of three planets including XO-4b. To better constrain the ephemerides of the planetary transits \citet{todorov2012} make additional $I_{c}$ band ground based primary transit observations using the Universidad de Monterrey Observatory (UDEM) 0.36-m telescope. We do not include secondary eclipse data in our data analysis, but we do include the UDEM $I_{c}$ band ground based primary transit observations in our global analysis, including the previously unpublished UT 2012-01-07 data, kindly provided by the authors.

We convert the times from the \citet{todorov2012} UDEM observations from $\mathrm{BJD_{UT}}$ to \bjdtdb~using the IDL code {\tt JDUTC2JDTDB} to maintain uniform time stamps across all available light curves. While this routine is intended to convert Geocentric Julian Date from UTC to TDB, it is accurate to 30 ms when using it to convert Barycentric Julian Dates instead \citep{east2010t}, which is more than sufficient for our purposes.

\section{Models}

\subsection{Transits and Radial Velocity}

The models used in this paper to fit the transit and orbital radial velocity data are unchanged from the original \citet{east2013} EXOFAST paper. The original EXOFAST paper ignores a number of effects, including RM and transit timing variations (TTVs), that we now include to maintain consistency with work done by the other groups.

\subsection{Rossiter-McLaughlin Effect}\label{rmsec}

We also consider the radial velocity data taken during transit, and as such must consider models for the RM effect \citep{rossiter24,mclaughlin24}. The precise model used to model the RM effect can have a significant effect on the inferred parameters, and yet this can vary from system to system \citep{jaj2008,hirano2010}.  A number of RM models exist, and we investigate two separate models here.

The ambiguity of the proper model of the RM effect that should be used comes from the fact that the RM effect is not, in fact, due to a change in the radial velocity of the star. The radial velocity measurement is made by estimating some measure of the ``centers'' of known absorption lines and their change in central wavelength relative to laboratory measurements. If the shape of the absorption features changes in such a way as to change the odd moments of the lines, then this can result in a change in the measured line centers, and as a result be attributed to a change in the radial velocity of the star. The precise relation between the change in the line shape and the inferred change in the line center will depend on many intrinsic properties of the star, such as its effective temperature, surface gravity, and rotation rate, as well as the precise algorithm used to estimate the line centers.

Thus the RM effect manifests as an anomalous radial velocity measurement $\Delta v_{\rm{RM}}$ made during the primary transit, where the anomalous signal is due to a change in the shape of the absorption feature rather than to the motion of the host star. The change in shape is due to the transiting planet preferentially blocking light emitted from the rotating host star. The blocked light, depending on the position of the planet as viewed from the observer, may be red or blue shifted relative to the center of the star due to the rotation of the host star. As this light is blocked it can introduce an asymmetry to the absorption feature \citep{gaudiwinn2007}. Because the anomalous signal is dependent on a change in the shape of the absorption feature, the measured $\Delta v_{\rm{RM}}$ will depend heavily on the method used to measure the radial velocity signal, and whether that method is sensitive to changes in the shape of the absorption features.

The RM anomalous radial velocity shift has a strong dependence on the host star's rotation $v\sin{I_{*}}$ and the orbit of the planet relative to the star's spin axis $\lambda$. The RM signal is also dependent on the radius of the planet in stellar radii $R_{p}/R_{*}=p$, the limb-darkening parameters $u_{1}$ and $u_{2}$, and the path the planet takes over star which can be described by the inclination $i$ and orbital distance $a$ or by the impact parameter $b$. \citet{winn2010} gives an approximation for the maximum amplitude of the RM effect as:
\begin{eqnarray}
\Delta v_{\rm{RM}}\approx p^{2}\sqrt{1-b^{2}}\left(v\sin{I_{*}}\right)
\end{eqnarray}
When the effect is observed it is possible to place constraints on the spin-orbit alignment $\lambda$, project rotational velocity $v\sin{I_{*}}$, and the impact parameter $b$. The transit light curves and previous RV studies provide us with independent constraints on $v\sin{I_{*}}$ and $b$, and these improve the constraint on $\lambda$.

There are two different RM models we investigate to place the constraint on $\lambda$, those based on the moment method \citep{ohta2005,ohta2009} (OTS) and those based on the cross-correlation method \citep{queloz2000,winn2005,narita2009,hirano2010}. There are other models available \citep{gimenez2006,boue2013}, but we focus on the OTS and cross-correlation methods as the default model in EXOFAST is that based on \citet{ohta2005} and previous work on XO-4b follows the cross-correlation method.

\subsubsection{Moment Method}

\citet{ohta2005} describe their derivation of the RM effect as an approximate but accurate analytic formula for the anomaly in the radial velocity curves. Their method approximates the velocity anomaly as the change in the first moment of the absorption line profile and uses only linear stellar limb-darkening. This work is followed by \citet{ohta2009} where the authors present theoretical predictions for the photometric and spectroscopic signatures of rings around transiting extrasolar planets that now include quadratic limb-darkening and terms for ringed extrasolar planets. These new expressions supersede the work in \citet{ohta2005}.

To express the radial velocity anomaly we first substitute the variable OTS used to represent stellar intensity to $\mathcal{I}_{*}$ to avoid confusion with the stellar inclination $\sin{I_{*}}$. OTS define the flux of the star as a function of the position of the planet
\begin{eqnarray}\label{fluxeq}
F=\frac{\iint \mathcal{I}(x,z)\mathrm{d}x\mathrm{d}z}{\iint \mathcal{I}_{*}(x,z)\mathrm{d}x\mathrm{d}z},
\end{eqnarray}
where x and z are the position of the center of the planet perpendicular to  and parallel to, respectively, the projected rotation axis of the star, and they integrate over the surface brightness of the unocculted stellar disk $\mathcal{I}_{*}$, and  $\mathcal{I}$ is the surface brightness of the occulted star, such that  $\mathcal{I}=0$ in the region occulted by the planet. OTS then formally express the radial velocity anomaly as the first moment of the relative flux
\begin{eqnarray}
\Delta v_{\rm{RM}}=-v\sin{I_{*}}\frac{\iint x\mathcal{I}(x,z)\mathrm{d}x\mathrm{d}z}{\iint \mathcal{I}_{*}(x,z)\mathrm{d}x\mathrm{d}z},
\end{eqnarray}
where they assume rigid rotation in the star, i.e. no differential rotation such that the projected velocity is constant along lines of constant $x$.

When we use these expressions we ignore the potential effects of planetary rings. Under this assumption, the expression can be reduced the following expression:
\begin{eqnarray}\label{eqoht}
\Delta v_{\rm{RM}}=x_{p}v\sin{I_{*}}F
\end{eqnarray}
where $x_{p}$ is the $x$ component of the planet's position in units of stellar radii and $F$ is the relative flux from Equation~(\ref{fluxeq}). The details of these expressions are included in \citet{ohta2009}, including detailed expressions for integrating the flux in Equation~(\ref{fluxeq}).

\subsubsection{Cross-Correlation Method}

An alternative method to measure the RM effect is to compare the anomalous radial velocity by cross-correlation with a stellar template spectrum. This method was used by \citet{queloz2000}, \citet{winn2005}, \citet{narita2009}, among others, and is described in detail by \citet{hirano2010}. This method involves creating an $N\times N$ element star, where $N^{2}\sim10^{6}$, and each element contains a solar spectrum. \citet{hirano2010} create a synthetic spectrum of the star by applying a stellar rotation kernel to redshift each element, multiply the spectral lines by the instrumental profile of their spectrograph, and integrate all of the elements minus those obscured by the transit to create simulated in-transit spectra. These simulated spectra are then fed through their analysis routine that fit the lines and adjust the shape of the model to best fit the data.

\citet{hirano2010} cover in detail the choice of absorption line profile (Gaussian, Voigt, etc.), and eventually settle on a form inspired by their Gaussian approximation:
\begin{eqnarray}\label{eqhir}
\Delta v_{\rm{RM}}=-Fv_{p}(p-qv_{p}^{2})
\end{eqnarray}
Here $F$ is the flux ratio and uses the same expression as that of Equation~(\ref{fluxeq}), and $v_{p}$ can be expressed as the position of the planet
\begin{eqnarray}
v_{p}(x,y)=\frac{\iint v(x',y')x'\sin{I_{*}}\mathrm{d}x'\mathrm{d}y'}{\iint \mathrm{d}x'\mathrm{d}y'}.
\end{eqnarray}
The parameters $p$ and $q$ are each functions of thermal broadening, micro-turbulent broadening, and the stellar rotation width that describe the shape of the line. The parameters $p$ and $q$ are empirically fit for each individual planetary system the method is applied to. For rigidly rotating stars $v(x,y)\sin{I_{*}}$ is a constant along lines of constant $x$ and the expression reduces to
\begin{eqnarray}
v_{p}(x,y)=xv\sin{I_{*}},
\end{eqnarray}
and reproduces the OTS
\begin{eqnarray}
\Delta v_{\rm{RM}}=-v\sin{I_{*}}\frac{\iint x'\mathcal{I}(x',z')\mathrm{d}x'\mathrm{d}z'}{\iint\mathcal{I}_{*}(x',z')\mathrm{d}x'\mathrm{d}z'}
\end{eqnarray}
result as noted by \citet{hirano2010}.

As a result, to first order, for which $p=1$ and $q=0$, Equation~(\ref{eqhir}) reduces to Equation~(\ref{eqoht}) and the two methods are identical.

Because $p$ and $q$ depend on precisely how the algorithm used to estimate the radial velocity measures the shape of the absorption profiles and  the line spread function of the spectrograph, it is possible to confuse a change in line shape as instrumental and not due to the RM effect. If ignored, this can lead to a misinterpretation of the RM effect, and as such this method must be applied to each system as outlined above to distinguish between the two effects. There are cases when the moment method and cross-correlation methods produce the same results \citep{winn2008,jaj2008}; however, for many cases the two methods result in statistically different inferences for the derived parameters of $v\sin{I_{*}}$ and $\lambda$. For reference, in the case of XO-4b, \citet{narita2010} derive and report $p=1.6159$ and $q=0.83778$.

\section{Data Analysis}\label{datasec}

To analyze the XO-4b data we use a custom version of the publicly available EXOFAST suite of IDL programs described in \citet{east2013}. This is a continuation of the same code used to fit the KELT discoveries \citep[e.g.,][]{siverd2012}, and includes the ability to simultaneously fit multiple transit light curves, both on the same telescope from different nights and from different telescopes, as well as the ability to include multiple radial velocity data sets. Additional modifications were made by us to update the default RM models, include alternative RM models, and to change how the estimated errors are scaled on the RM data.

We adopt Gaussian priors from MC08 on the stellar parameters $T_{\rm{eff}}$, $\log{g_*}$, \feh, and $v\sin{I_{*}}$, and use the reported values of $p=1.6159$ and $q=0.83778$ from N10 when using the \citet{hirano2010} based RM models. We do ignore secondary eclipse data from \citet{todorov2012} and assume $e=0$ as this is consistent with all of the other groups \citep{MC08,narita2010,todorov2012}.

\subsection{Narita Data vs Narita 2010}\label{compsec}
As a confirmation of our light curve analysis we make many of the same assumptions used in the N10 paper and compare the output from EXOFAST using the N10 data to the published results of the N10 paper. It should be noted that N10 include the same Gaussian priors we adopt, and additionally include a Gaussian prior on the period and $T_{0}$ from MC08. N10 also fix the limb-darkening parameter $u_{1}$ and treat the second limb-darkening parameter $u_{2}$ as a free parameter. In EXOFAST both $u_{1}$ and $u_{2}$ are calculated from $\log{g_*}$, $T_{\rm{eff}}$, \feh, and the observed band-pass from \citet{claret2011}. During this exercise we do not create a new free parameter for $u_{2}$, nor do we fix $u_{1}$, but we do include the Gaussian priors on period and $T_{0}$. Thus our results are not precisely comparable, although we expect any differences to be relatively minor.

Table~\ref{tabnaritacomp} contains the same parameters published in N10 (Narita$_{\chi^{2}}$) along with those produced by EXOFAST (EXOFAST$_{\rm{MCMC}}$). The values are consistent within the errors for all parameters. It should be noted that N10 use the minima in $\chi^{2}$ and $\Delta\chi^{2}=1$ to quote their best value and errors, where we typically quote the median values of the parameters and the 68\% confidence intervals from the MCMC chains. In order to more precisely compare our results to those of N10, we also fit a multi-dimensional hyperboloid to the output MCMC chains in order to infer the values of the parameters with the minimum $\chi^{2}$ and the values where $\Delta\chi^{2}=1$ (EXOFAST$_{\chi^{2}}$). 

As we did not create a new free parameter for $u_{2}$, the constraint on the second limb-darkening parameter is calculated from the parameters $\log{g_*}$, $T_{\rm{eff}}$, \feh, and the band. Thus the uncertainty on $u_{2}$ is simply a reflection on the covariance of $u_{2}$ with these parameters. We generally find values and uncertainties that are in good agreement with those of N10, verifying that the methods used in N10 are comparable to those used in EXOFAST. The final column of Table~\ref{tabnaritacomp} shows the differences in the two minimum $\chi^{2}$ values divided by the uncertainties in quadrature for reference. The differences between methods are less than $1~\sigma$ and any differences may be attributable to our slightly different methods of estimating the values for the minimum $\chi^{2}$ and $\Delta\chi^{2}=1$.

\begin{deluxetable*}{lcccc}[t]
\centering
\tablecaption{A comparison of XO-4b median values and 68\% confidence intervals from the EXOFAST MCMC chains as compared to the $\chi^{2}$ minimum and $\Delta\chi^{2}=1$ values from EXOFAST and the N10 published values.}
\tablehead{\colhead{Parameter} & \colhead{EXOFAST$_{\rm{MCMC}}$} & \colhead{EXOFAST$_{\chi^{2}}$} & \colhead{Narita$_{\chi^{2}}$} & \colhead{$\vert\Delta/\sqrt{\sigma^{2}+\sigma^{2}}\vert$\footnotemark[1]}}\\
\startdata
$P$ [days]\dotfill & $4.1250801\pm0.0000028$ & $4.1250801\pm0.0000028$ & $4.1250828\pm0.0000040$ & $0.55$ \\%$-0.55$ \\
$T_C$ [\bjdtdb]\dotfill & $2454485.93346\pm0.00023$ & $2454485.93346\pm0.00023$ & $2454485.93323\pm0.00039$ & $0.51$ \\
$K$ [m/s]\dotfill & $172\pm11$ & $174\pm11$ & $168.6\pm6.2$ & $0.43$\\
$v\sin{I_*}$ [m/s]\dotfill & $8660_{-450}^{+470}$ & $8820\pm530$ & $8900\pm500$ & $0.11$ \\%$-0.11$ \\
$\lambda$ [degrees]\dotfill & $-41.2_{-7.5}^{+9.2}$ & $-47.0\pm14$ & $-46.7_{-6.1}^{+8.1}$ & $0.02$ \\%$-0.02$ \\
$a/R_*$\dotfill & $7.58_{-0.19}^{+0.12}$ & $7.73\pm0.23$ & $7.68\pm0.11$ & $0.20$ \\
$R_{P}/R_{*}$\dotfill & $0.08770_{-0.00052}^{+0.00055}$ & $0.08739\pm0.00060$ & $0.0881\pm0.0007$ & $0.77$\\%$-0.77$\\
$i$ [degrees]\dotfill & $88.30_{-0.73}^{+0.60}$ & $89.01\pm0.87$ & $88.8\pm0.6$ & $0.20$\\
$u_{2}$\dotfill & $0.3021\pm0.0022$ & $0.3098\pm0.0027$ & $0.35\pm0.11$ & $0.37$ \\%$$-0.37$ \\
$\gamma_{1}$ [m/s]\dotfill & $-3.3\pm8.1$ & $ -4.3\pm 8.4$ & $-0.1\pm2.9$ & $0.47$\\%$-0.47$\\
\enddata
\label{tabnaritacomp}
\footnotetext[1]{This is defined as $\vert(\rm{EXOFAST}_{\chi^{2}}-\rm{Narita}_{\chi^{2}})/\sqrt{\sigma_{\rm{EXOFAST}}^{2}+\sigma_{\rm{Narita}}^{2}}\vert$}
\end{deluxetable*}

\subsection{DEMONEX Data}
We converted all times in the DEMONEX data to barycentric Julian date in the barycentric dynamical time (\bjdtdb) as advocated in \citet{east2010t}. We performed standard data reduction procedures for the raw DEMONEX XO-4 data. These include bias correction, dark subtraction, flat fielding, and due to our Sloan $z$ band observations we additionally perform fringe corrections. To perform the fringe corrections a master fringe image is created by taking the median of the nearest 100 images separated by a minimum time step, where the stars are masked out in each image. The background and amplitude of the master fringe image are fit and then subtracted off of each science image.

We use AstroImageJ\footnote{\emph{http://www.astro.louisville.edu/software/astroimagej/}} (AIJ) to perform aperture photometry on the target star and a set of comparison stars to create normalized relative photometric light-curves for XO-4. We use an initial set of comparison stars based on their similar counts to XO-4. For each night the flux from the target star XO-4 was divided by the sum of the flux from the set of comparison stars in the image and normalized to unity. We then select the the comparison star that gives the lowest out-of-transit (pre-ingress and post-egress) root-mean-square (RMS) for XO-4, and we add additional comparison stars only if their inclusion reduces the out-of-transit RMS of the normalized light curve from XO-4. This process is repeated for each night to create the lowest RMS light curves possible and the final light curves are shown in Table~\ref{demonexdata}.

\begin{deluxetable*}{ccccc}[t]
\centering
\tablecaption{DEMONEX sample data including trend parameters for XO-4b.}
\footnotetext{\textbf{Notes}. This table is available in its entirety in the online journal. A portion is shown here for guidance regarding its form and content.}
\tablehead{\colhead{Time[\bjdtdb]} & Normalized Flux & Flux Error & x & y}
\startdata
\sidehead{UT 2008-11-11:}
2454782.76202800 & 1.001644000 & 0.00239300000 &  866.200393 & 1219.753505 \\
2454782.76282000 & 1.008380000 & 0.00210400000 &  870.227406 & 1220.639406 \\
2454782.76361400 & 1.016223000 & 0.00197900000 &  873.242392 & 1220.334975 \\
2454782.76440700 & 1.004863000 & 0.00180800000 &  875.858008 & 1220.666278 \\
2454782.76520100 & 1.005626000 & 0.00175800000 &  877.882314 & 1219.812195 \\
2454782.76599400 & 1.005393000 & 0.00180700000 &  879.948887 & 1218.680215 \\
2454782.76678600 & 1.010737000 & 0.00178400000 &  881.385388 & 1218.341993 \\
2454782.76757900 & 1.006882000 & 0.00173300000 &  881.728714 & 1217.411449 \\
2454782.76837200 & 1.002178000 & 0.00176400000 &  883.499507 & 1216.778263 \\
2454782.76916700 & 1.007074000 & 0.00174700000 &  883.487292 & 1216.278131 \\
\enddata
\label{demonexdata}
\end{deluxetable*}

\begin{figure}[t]
\centering
\includegraphics[width=8cm]{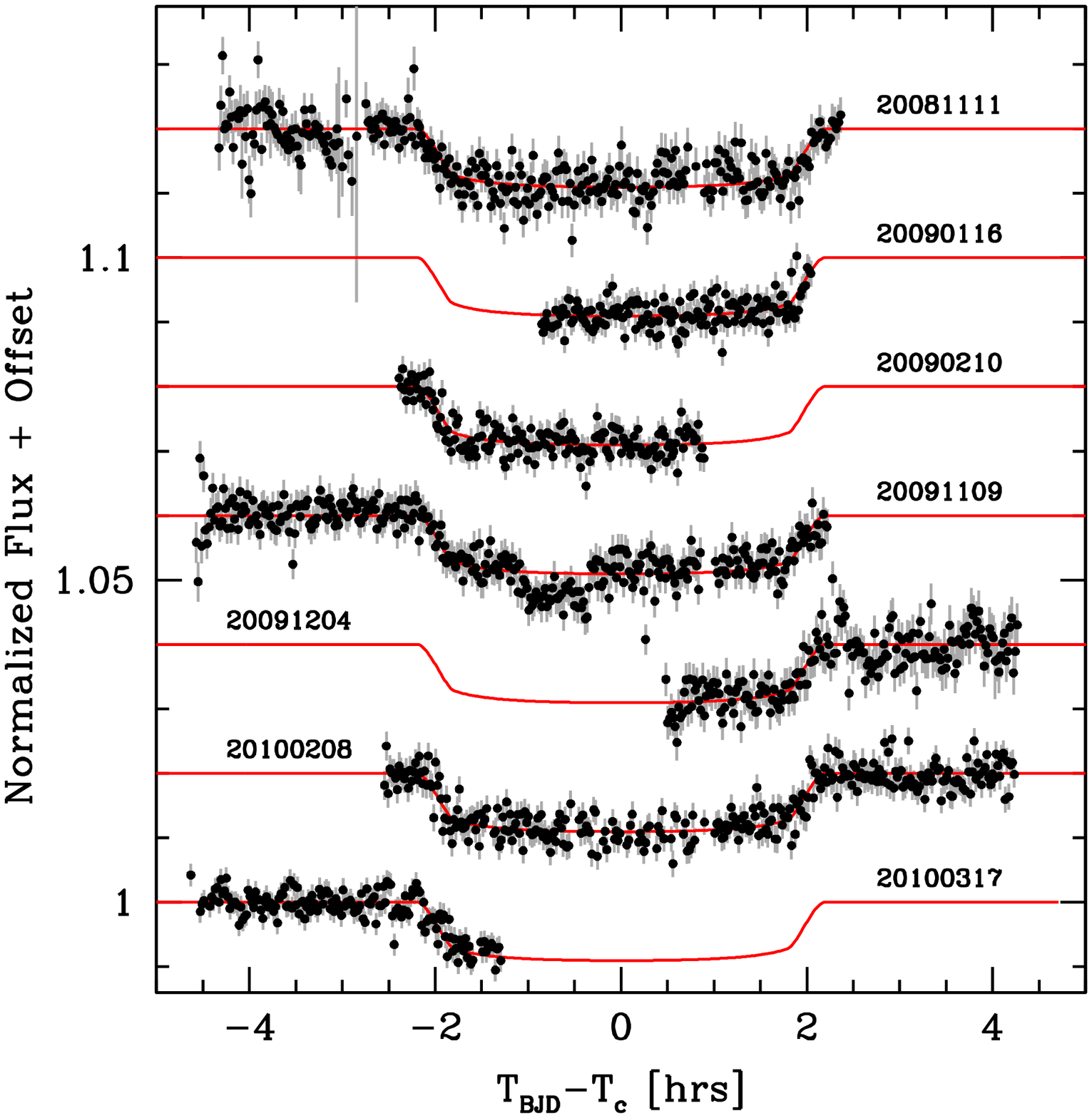}
\centering
\includegraphics[width=8cm]{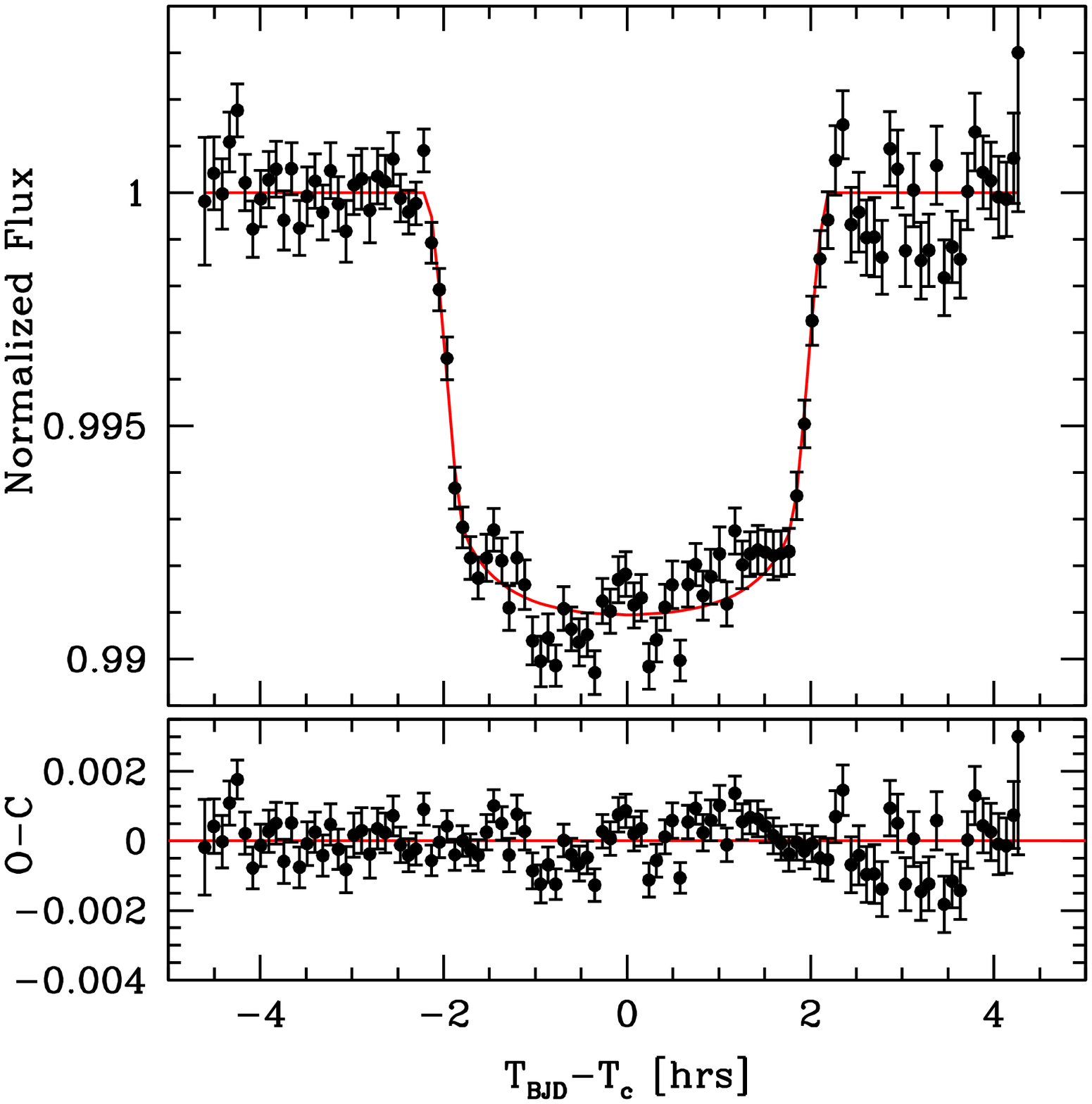}
\caption{\label{demonextot}Top: Each of the 7 DEMONEX light curves with and arbitrary offset fitted using only DEMONEX data and detrended against the $x$ and $y$ position of the star on the chip, and the time (\bjdtdb). The error bars are plotted in gray and the best-fit model is over plotted in red.\\\\
Bottom: The binned data and the best-fit model are shown at the bottom with residuals. Binned data is not used in the analysis but shown to better display the overall quality of the data and the statistical power of the DEMONEX data.}
\end{figure}

Significant trends in the data are immediately apparent. These were mostly due to issues with the mount's ability to track and guide and so the trends were correlated with XO-4b's $x$ and $y$ location on the chip. After detrending the data, the 7 individual light curves are shown in Figure~\ref{demonextot} with the binned data and residuals below. A summary of the planetary parameters derived from the transits are shown in Table~\ref{demonextab}.

\begin{deluxetable}{lcc}[t]
\tablecaption{Median values and 68\% confidence interval for Primary Transit Parameters of XO-4b new DEMONEX observations.}
\tablehead{\colhead{~~~Parameter [Units]} & \colhead{Value}}
\startdata
\sidehead{Primary Transit Parameters:}
~~~$R_{P}/R_{*}$\dotfill & $0.0893\pm0.0012$\\
           ~~~$a/R_*$\dotfill & $7.70_{-0.21}^{+0.11}$\\
                          ~~~$i$[degrees]\dotfill & $88.47_{-0.81}^{+0.58}$\\
                               ~~~$b$\dotfill & $0.205_{-0.076}^{+0.10}$\\
                             ~~~$\delta$\dotfill & $0.00797_{-0.00021}^{+0.00022}$\\
                    ~~~$T_{FWHM}$[days]\dotfill & $0.1672\pm0.0011$\\
              ~~~$\tau$[days]\dotfill & $0.01569_{-0.00050}^{+0.00096}$\\
                     ~~~$T_{14}$[days]\dotfill & $0.1831_{-0.0014}^{+0.0016}$\\
   ~~~$P_{T}$\dotfill & $0.1183_{-0.0017}^{+0.0033}$\\
              ~~~$P_{T,G}$\dotfill & $0.1414_{-0.0021}^{+0.0040}$\\
                ~~~$u_{1Sloanz}$\dotfill & $0.1745\pm0.0059$\\
             ~~~$u_{2Sloanz}$\dotfill & $0.3019_{-0.0022}^{+0.0021}$\\
\enddata
\label{demonextab}
\end{deluxetable}

\subsubsection{Detrend Parameters}\label{detrendsec}

EXOFAST can take trends in the photometric data and remove them to improve the quality of data. We investigate which, if any, detrend parameters in the DEMONEX data are significantly affecting the data quality and thus should be removed. We consider the the position on the star on the CCD, $x$ and $y$, the time (\bjdtdb), and airmass $\sec{z}$ as these are typical detrend parameters used in other studies \citep[e.g.,][]{collins2014}. We take the case of no detrending and compare the results when the light curves were detrended against a single parameter. We then began adding in additional detrend parameters to detrend simultaneously. We notice that there are significant trends in all four detrend parameters when fitted individually. Once all of the parameters are included, the derived parameters converge to within $1~\sigma$ of values previously published in the literature as well as the values derived when we use only the N10 data in EXOFAST from \S\ref{compsec}. This suggests that including all available detrend parameters improves our analysis, at the expense of more computational time and creating the potential of a new local minima in the $\chi^{2}$ space that we search through. As airmass and time are correlated we only include time to reduce the parameter space that the EXOFAST MCMC chains must search through and to speed up the computational time. Thus our final DEMONEX data set is detrended against $x$, $y$, and \bjdtdb.

\begin{figure}[t]
\centering
\includegraphics[width=8cm]{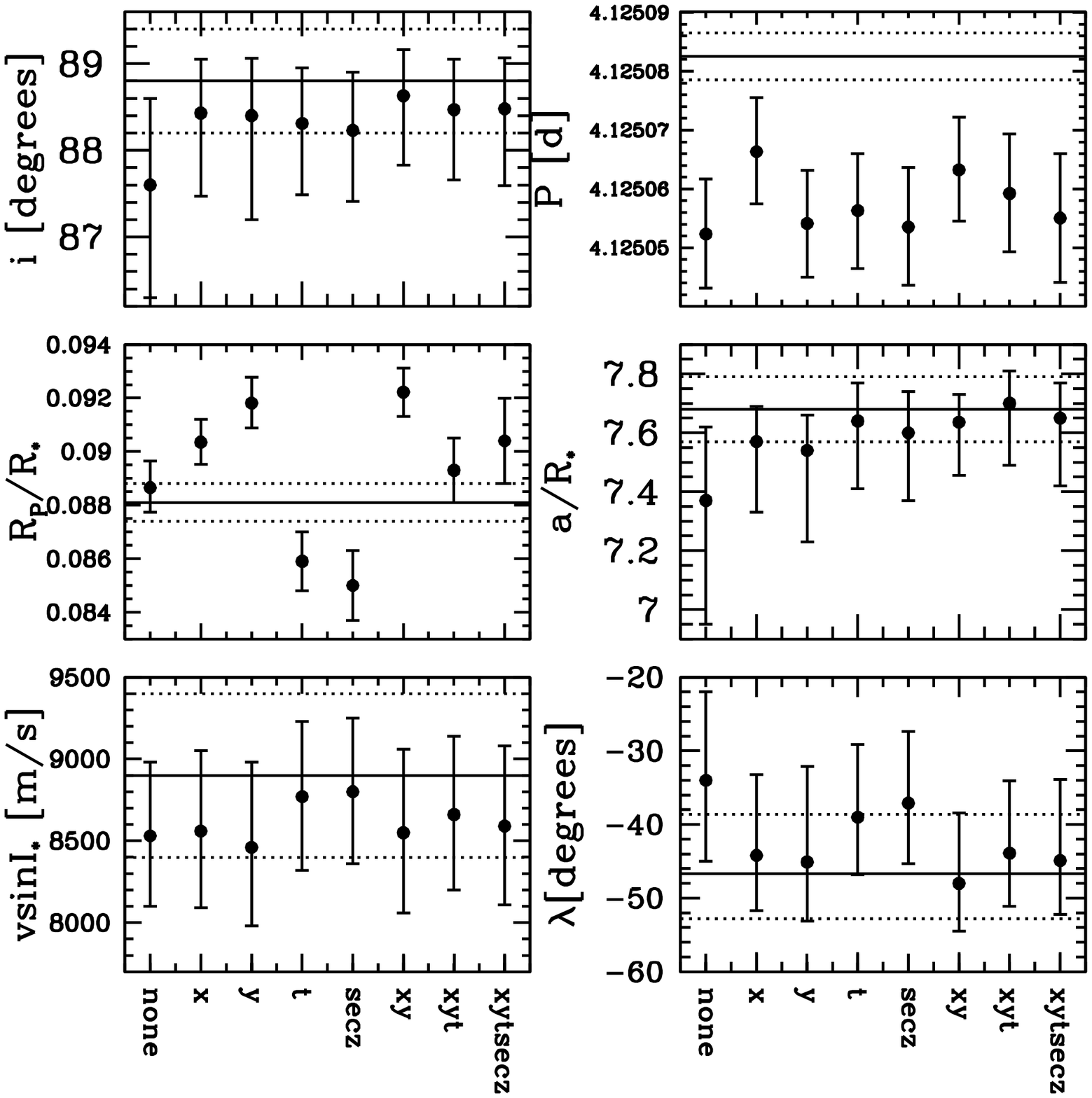}
\caption{\label{detrend1} Various derived parameters in the DEMONEX light curves as a function of the chosen detrend parameters: the position of the star on the CCD ($x$, $y$), time $t$, and airmass $\sec{z}$. The x-axis containing the detrend parameters is the same for each plot and lists which detrend parameters were used i.e. no detrending, detrending in $x$ only, $y$ only, etc. including detrending against multiple parameters simultaneously. The solid and dashed lined represent the published Narita results and their error bars. In most cases the inferred value is still consistent, i.e. within $1~\sigma$, with previous results when no detrending parameters are chosen, but the answers tend to converge to more consistent values when a larger number of detrend parameters are included. It should be noted that for the parameter $R_{P}/R_{*}$ the change in the detrend parameter can change the inferred value by over 3~$\sigma$ and the systematic uncertainty due to the selection of detrend parameters is far greater than the statistical uncertainty estimated from the data itself. We did not apply an additional Gaussian prior on $T_{c}$ (see \S\ref{compsec}) which explains the difference in our inferred value for the period relative to the published Narita value.}
\end{figure}

The parameter most sensitive to the quality of data and selection of detrend parameters is $R_{P}/R_{*}$, while others (such as $v\sin{I_{*}}$) are weakly dependent on the detrend parameters. However many of the derived parameters are correlated with $R_{P}/R_{*}$, parameters such as the spin-orbit alignment $\lambda$ that depend on both $R_{P}/R_{*}$ and $v\sin{I_{*}}$, are still subject to the choice of included detrend parameters. Again we notice that most parameters converge to previous estimates as more detrend parameters are included as seen in Figure~\ref{detrend1}. It it worth noting that for the parameter $R_{P}/R_{*}$ the change in the inferred value for $R_{P}/R_{*}$ can vary by over 3~$\sigma$ depending on the choice of, and number of, detrend parameters. Thus the systematic uncertainty due to the selection of detrend parameters is far greater than the statistical uncertainty quoted and derived from the data itself.

\begin{figure}[t]
\centering
\includegraphics[width=8cm]{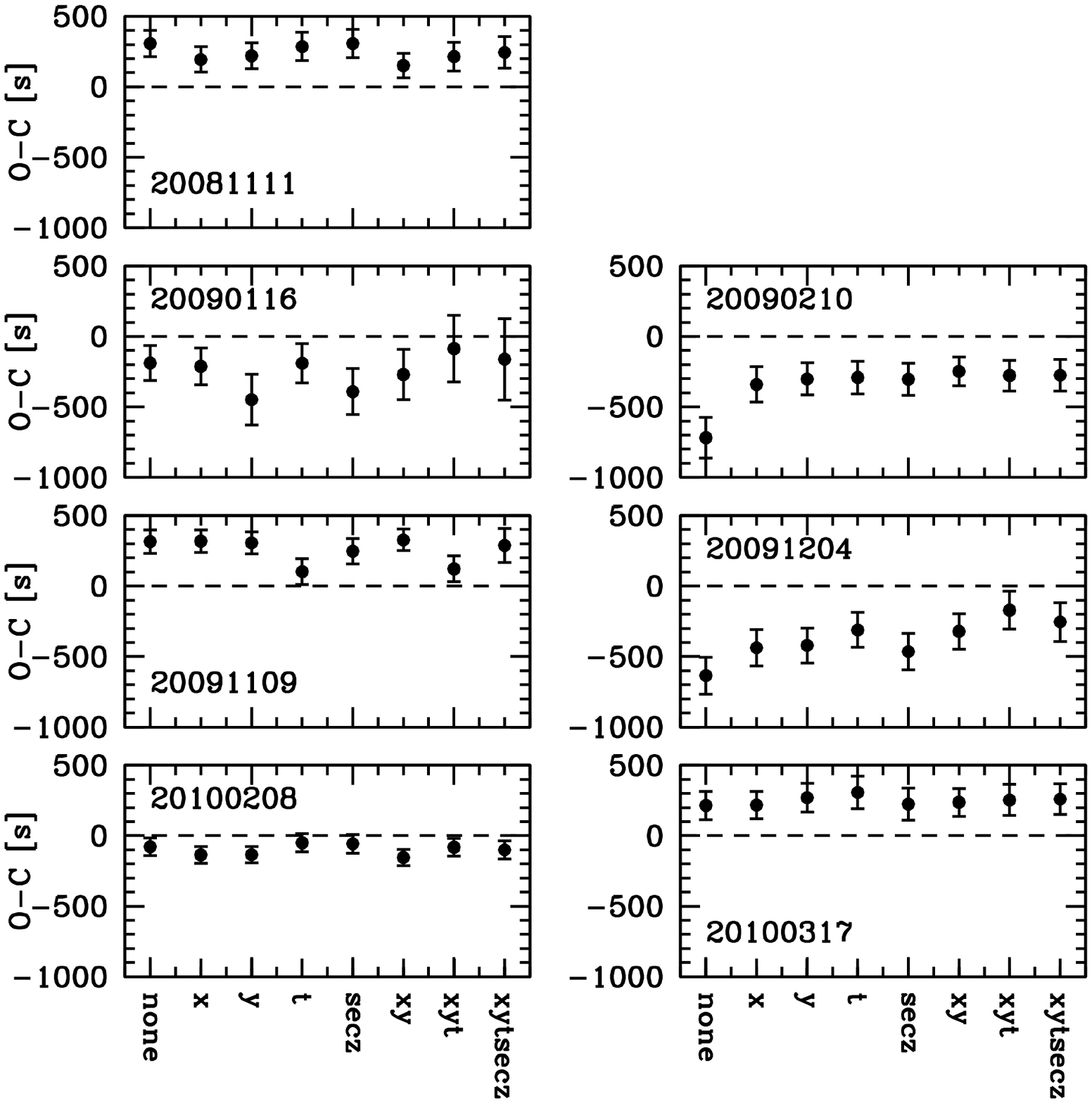}
\caption{\label{ttv2}Various derived offsets from the calculated central transit time to the measured $T_{c}$ as a function of the various combinations of detrending parameters used for the 7 DEMONEX nights. For some nights the changes in O-C are consistent within the errors, while for others the inferred O-C is dependent on the choice of detrend parameters. It should be noted that most of the inferred O-C values are still inconsistent with zero and suggest the presence of a transit timing variation independent of the choice of detrend parameter. We believe that these TTVs are unlikely to be real, but rather are the result of as-yet unrecognized systematics.}
\end{figure}

Additionally we look at the effects of detrending on the inferred TTVs by enabling the time of the central transit to be a free parameter, as shown in Figure~\ref{ttv2}. We should note that not all of the possible combinations of detrend variables were able to be selected as the increased freedom of floating central transit times and the detrending parameters resulted in many chains failing to meet the convergence criteria. To counteract this a Gaussian prior was placed on the period to allow the chains to converge. This prior is only used when detrending DEMONEX data only. In some cases the inferred variations in the derived O-C is consistent within the uncertainty regardless of which detrending variable is selected. There are still cases where the O-C varies at the few $\sigma$ level, indicating that the choice of detrend variable once again matters and the presence of systematic uncertainties that exist in addition to the quoted statistical uncertainties. It is still the case that even when using detrending parameters, multiple nights have an O-C that imply a significant TTV. 

It is not clear to the authors which detrend parameters, or combinations of detrend parameters, are the ``correct'' parameters to detrend against. Including detrend parameters has significant implications on the derived parameters, especially $R_{P}/R_{*}$, but we do not yet have an objective way to determine which set of parameters is ``correct''. We consider some metric involving $\chi^{2}$ minimization, but note that this assumes uncorrelated data which we know not to be the case. We understand that our data contain systematics, and detrending is one such effort to reduce the effects of these systematics. We also note that including all of the available parameters produces inferred values consistent with previous measurements, but this can vary at the $3~\sigma$ level to the inferred value when detrend parameters are not used. We therefore conclude that detrended data inherently contain additional systematic uncertainties, typically not quoted, and detrending must be applied carefully in order to avoid biasing the inferred parameters. Additionally readers should be cautious when trusting the inferred values of these parameters in this and other papers. We have made efforts to quantify and reduce these systematic uncertainties, but do not yet have an ideal way in which to eliminate them altogether. Our study of these effects is however a step in the right direction.

\subsection{Combined Data Set}

\begin{figure}[t]
\centering
\includegraphics[width=8cm]{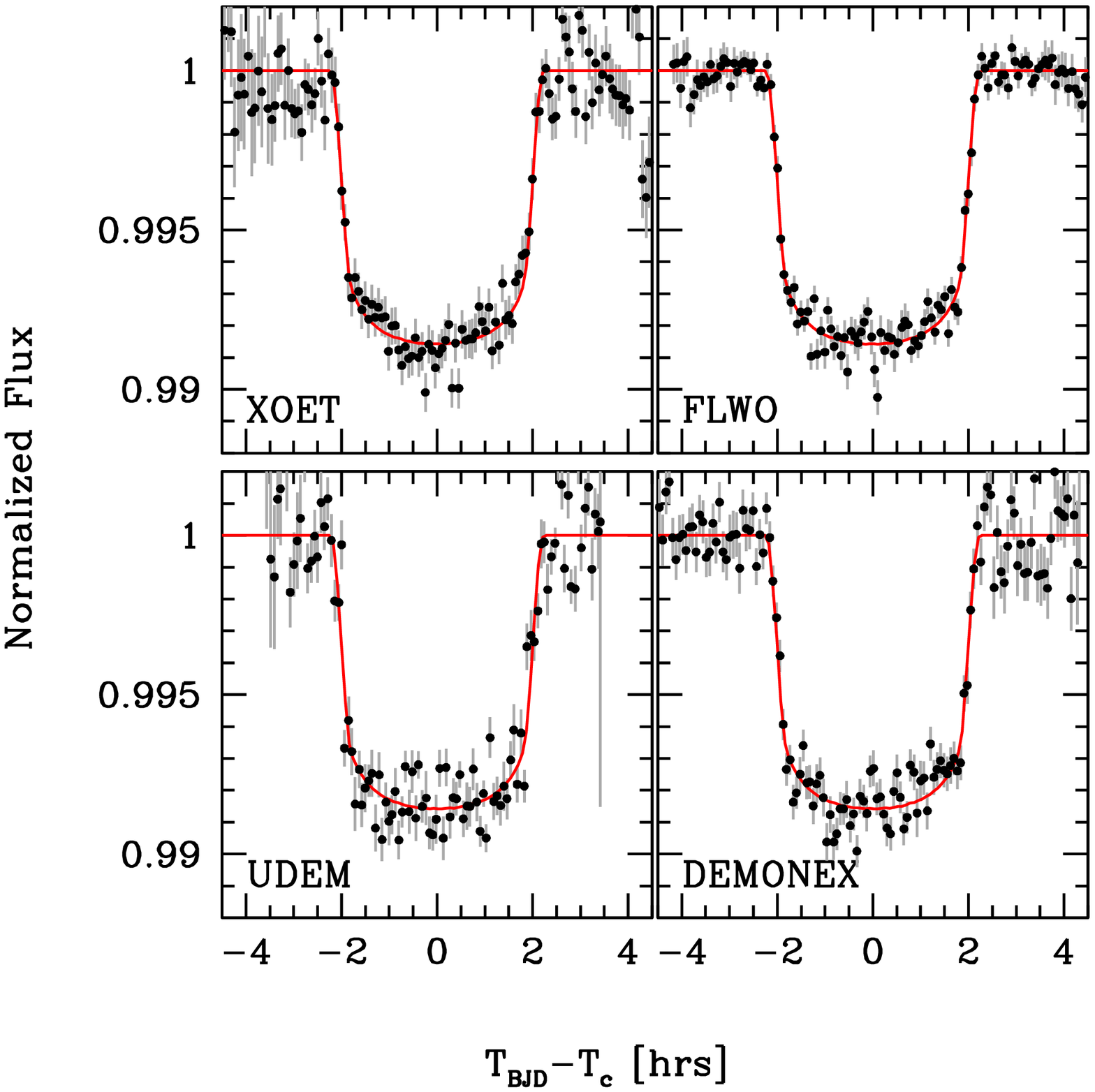}
\caption{\label{lc0} Binned light curves from each telescope to show relative quality of data from each data source. The best-fit global model is plotted in red. Each set of data is binned in the same way. The DEMONEX data is detrended against $x$, $y$, and $t$ while all others are detrended against $t$. Binned data is not used in the analysis but shown to better display the overall quality of the data.}
\end{figure}

\begin{figure}[t]
\centering
\includegraphics[width=8cm]{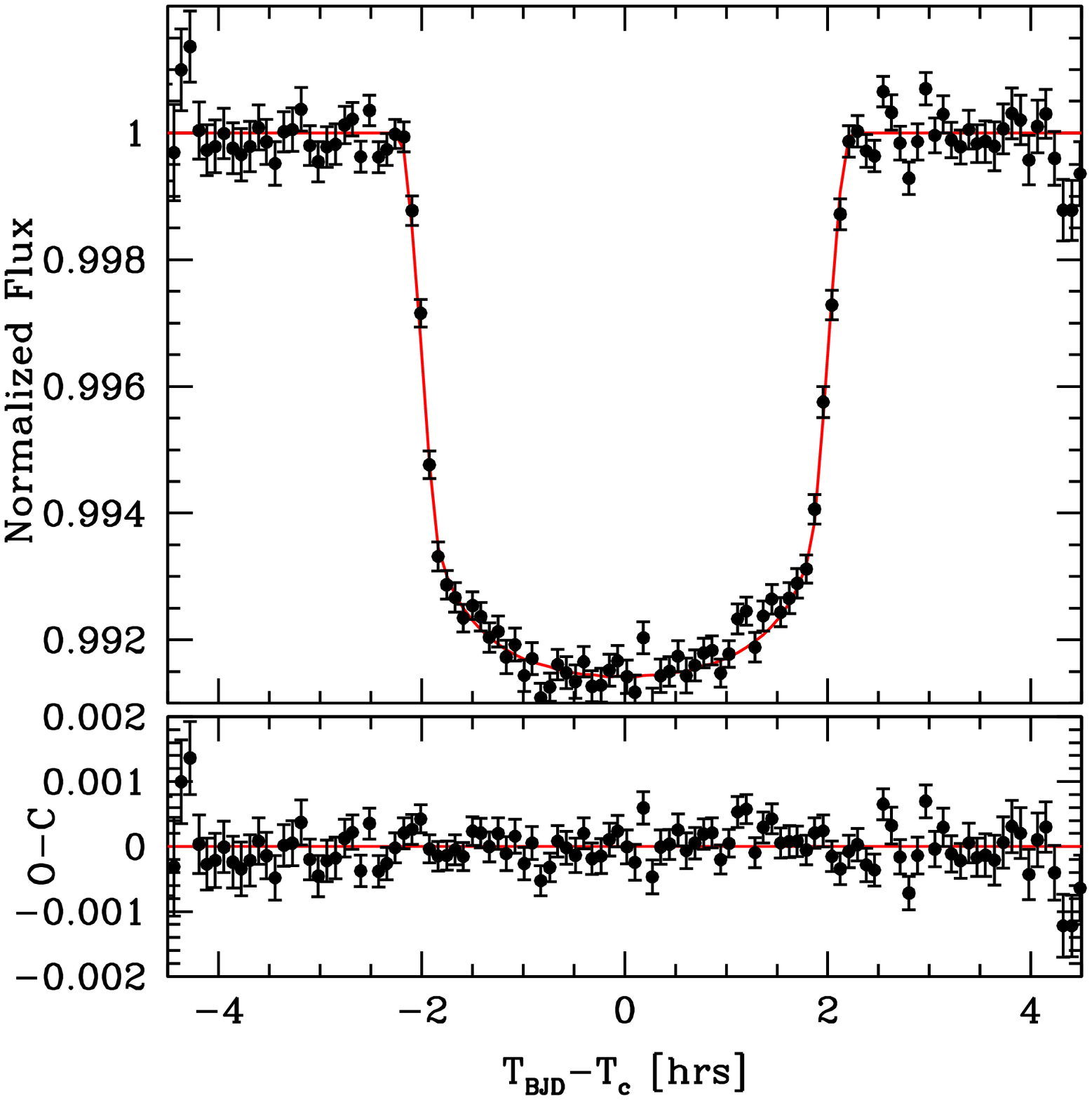}
\caption{\label{lc} All of the data globally fit by EXOFAST from the four different telescopes (XOET having 9 nights, FLWO 4 nights, UDEM 4 nights, and DEMONEX having 7 nights), phased and binned. Over-plotted in red is the best fit model for the global analysis and the residuals of the binned data are shown at the bottom. Binned data is not used in the analysis but shown to better display the overall quality of the data and the statistical power of the total data set.}
\end{figure}

Having demonstrated that the DEMONEX data (when properly detrended) are consistent with the results from previous analyses, we now combine the light curve data from the detrended DEMONEX data, MC08 XO Extended Team data (XOET), N10 data (FLWO), and \citet{todorov2012} optical data (UDEM) to create our combined primary transit data set. Additionally we detrend each of the non-DEMONEX data sets against time as we noticed trends in the XOET data. This final data set includes 24 primary transit light curves of XO-4b, covering 21 different nights, in 5 different bands. There are 7 transits taken from the 0.5-m DEMONEX telescope, 4 transits taken with the FLWO telescope from the N10 data set, 9 transits from the XO Extended Team MC08 data set, and 4 transits from the UDEM telescope in the \citet{todorov2012} data set. These are shown binned by telescope in Figure~\ref{lc0} and binned together in Figure~\ref{lc}. Combined there are 17 full transits containing both ingress and egress with multiple transits covered in multiple wavelengths or by different telescopes. The remaining light curves include 7 partials with 4 containing only the ingress and 3 containing only the egress.


\subsubsection{Data Diagnostics}
\begin{figure}[t]
\centering
\includegraphics[width=8cm]{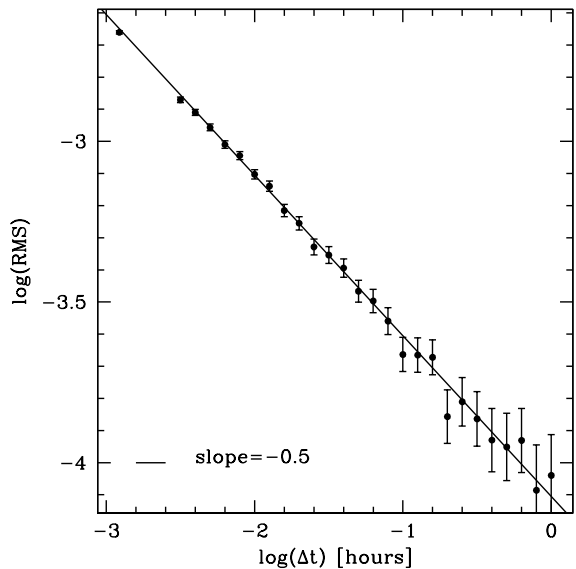}
\caption{\label{av} The RMS of the residuals of the individual data points from all of the light curves at various bin sizes in time. The weighted RMS of the unbinned data is 0.00221 as seen in the data point to the top left. We find that the RMS decreases as $\sqrt{t}$ as expected in the photon-noise limited regime if the data points are uncorrelated. This decreases to $0.000276$ when binned on $5~$min intervals as shown in Figure~\ref{lc}.}
\end{figure}

To look at the overall quality of our data, we look at the RMS of the residuals of the combined light curves shown in the bottom inset of Figure~\ref{lc}. We find that the weighted RMS of the $\sim$6700 data points within 4~hours of the central transit to have a fractional RMS in the residuals of the normalized flux of $0.00221$, or a factor of 3.4 smaller than the transit depth. To test the quality of our data we also verify that the RMS decreases as $\sqrt{t}$ when the data are binned. We provide an Allan variance plot, shown in Figure~\ref{av}, where we increase the bin-size of the combined data set light curves and verify that binning the data at larger intervals does in fact decrease the weighted RMS as $\sqrt{t}$ as expected if the errors are uncorrelated. When the data are binned at $5~$min intervals, as shown in Figure~\ref{lc}, the fractional RMS in the residuals of the normalized flux decreases to $0.000276$, or a factor of 27 smaller than transit depth.

\begin{figure}[t]
\centering
\includegraphics[width=8cm]{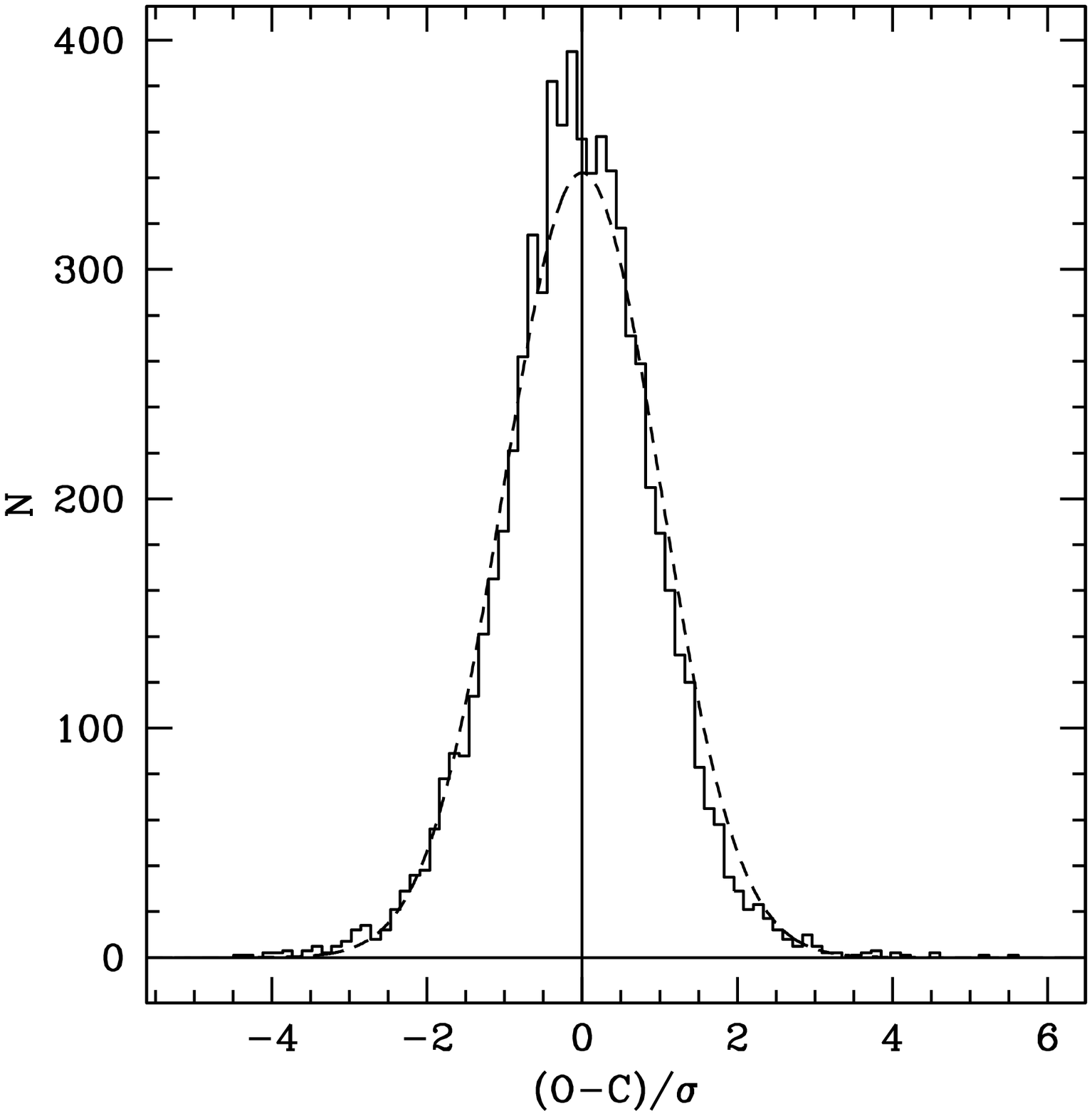}
\caption{\label{hist} The distribution of the residuals of the individual data points from all of the light curves normalized by their uncertainties. We note that we have scaled the uncertainties in the individual light curves by a constant multiplicative factor in order to force the $\chi^{2}/dof\sim1$. We find that the distribution is not perfectly Gaussian, with a larger number of points with values of (O-C)/$\sigma$ near zero than expected.}
\end{figure}

The most natural interpretation of the uncertainties we quote on the derived parameters assumes that the photometric uncertainties in the data are both uncorrelated and Gaussian distributed. In Figure~\ref{hist} we plot the distribution of the residuals in the light curves from all of the data sets normalized by their uncertainties. We find that the distribution is not perfectly Gaussian, with a larger number of points with values of (O-C)/$\sigma$ near zero than expected. This implies that our process of scaling the uncertainties by a constant factor is not entirely capturing the true nature of the systematic errors.

\subsubsection{Mass-Radius}

In addition we test the effects of using two methods to resolve the mass-radius degeneracy. The default relation for EXOFAST is based on the Torres relation, while the updated EXOFAST uses the Yale-Yonsei (YY2) isochrones. The Torres relation is based on \citet{torres2010} who provide empirical estimates of stars with precise mass and radius measurements to derive simple polynomial functions of $T_{\rm{eff}}$, $\log{g_*}$, and \feh~that yield $M_{*}$ and $R_{*}$ with scatter within the relations of 6\% and 3\%, respectively. As an alternative, we also use isochrones based on \citet{YY1} and \citet{YY2} which provide sets of isochrones over a wide range of metallicities and ages scaled to the solar mixture. The update to the EXOFAST code and implementation is described in \citet{east2015}.

We use both relations in our analysis and find that the constraints are tighter for $M_{*}$ and $R_{*}$ when using the YY2 isochrones. In the cases of the mass and radius of the star (and therefore the planet) the constraints are tighter by 50-95\%. As a result we use the results from the YY2 over those derived from the Torres relation, but it is clear that the two methods are in very good agreement.

\subsubsection{Spin-Orbit $\lambda$}

\begin{figure}[t]
\centering
\includegraphics[width=8cm]{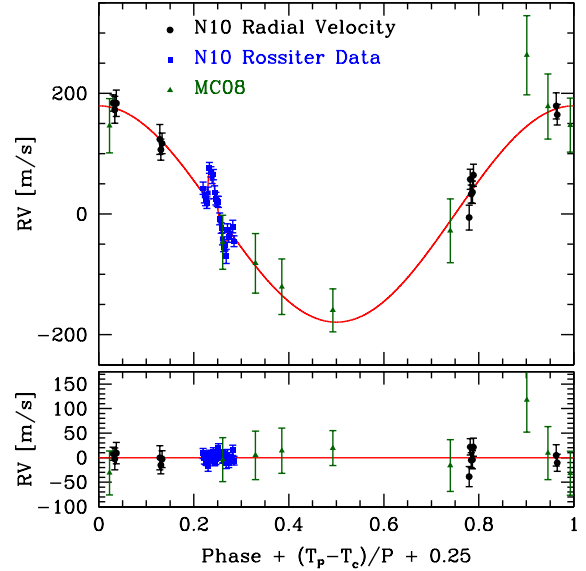}
\caption{\label{rv}Phased radial velocity curves for the two data sets used. The best fit model is over plotted in red with residuals below. The N10 data set is split into data during transit (N10 RM) and data out of transit (N10 RV). This is done to get a more accurate estimate of the error scaling on the Rossiter-McLaughlin data.}
\end{figure}

\begin{figure}[t]
\centering
\includegraphics[width=8cm]{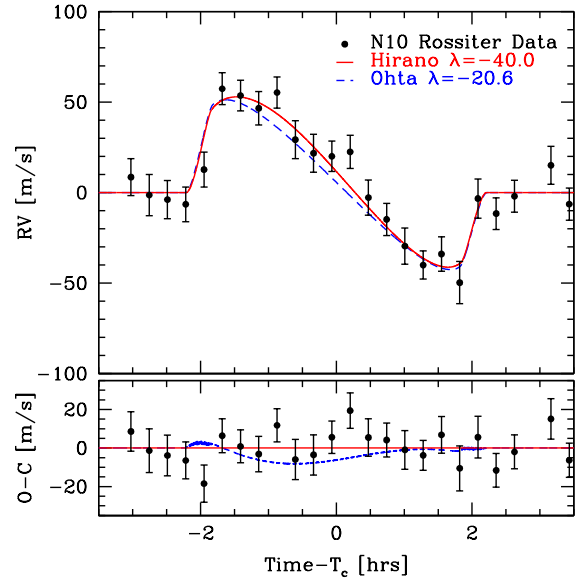}
\caption{\label{rm}The N10 Rossiter-McLaughlin data set and the two best fit models. Residuals relative to the Hirano model are shown below. The two models fit the data with two different values of $\lambda$ that vary at the 1-$\sigma$ level. There is a $\Delta\chi^{2}_{\mathrm{RM}}=2.5$ between the two RM data sets using the two RM models with the Hirano model having the lower $\chi^{2}$ value.}
\end{figure}

We simultaneously fit radial velocity data taken by MC08 and N10. The N10 radial velocity measurements also contain RM measurements taken during the transit and the RM data are fit as well. We separate the RM data to fit independent zero points and to scale the errors of the two data sets independently as we do not expect the night-to-night stellar variability and the stellar jitter to have the same magnitude and have the N10 RV and N10 RM data sets. The radial velocity and RM fits are shown in Figure~\ref{rv} and Figure~\ref{rm} respectively. Because we scale the individual data sets, we find that both models have a similar global fit with only a $\Delta\chi^{2}=0.6$ between the two, with the Hirano model having the lower $\chi^{2}$ value. However if we look at only the RM data and we use the Hirano scaled error bars on both data sets, we find that there is a $\Delta\chi^{2}_{\mathrm{RM}}=2.5$, again with the Hirano model having the lower $\chi^{2}$ value.


We find a significant difference between the inferred value of the spin-orbit misalignment angle $\lambda$ for the two RM models we consider. As shown in Figure~\ref{rm}, both models fit the data, but do so with different values of $\lambda$ and $v\sin{I_{*}}$. The two values derived for the spin-orbit alignment, $\lambda_{\rm{Hirano}}=-40.0_{-7.5}^{+8.8}$ and $\lambda_{\rm{OTS}}=-20.6_{-8.0}^{+9.0}$, disagree at the $1\sigma$ level. There is evidence that the OTS method can systematically miscalculate the amplitude of the anomalous radial velocity measurement $\Delta v_{\rm{RM}}$, but that the method still correctly recovers $\lambda$ \citep{benomar2014}.

\begin{figure}[t]
\centering
\includegraphics[width=8cm]{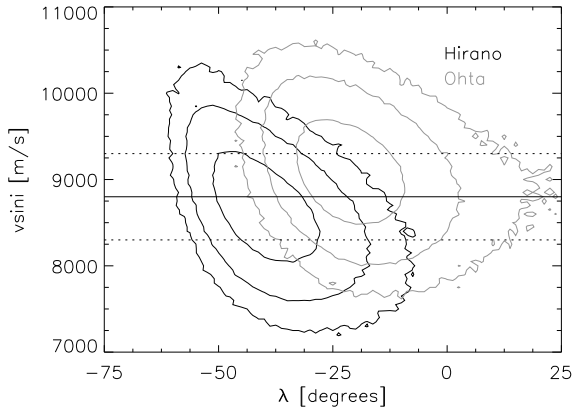}
\caption{\label{vsinilambda} 68, 95, and 99\% contours for the $v\sin{I_{*}}$ and $\lambda$ for the two Rossiter-McLaughlin models derived using the YY2 isochrones. These are equivalent to 1, 2, and 3$\sigma$ error contours. Over-plotted is the stellar prior on $v\sin{I_{*}}$ from \citet{MC08}. The two models prefer values of $\lambda$ that differ at the 1$\sigma$ level. The Ohta model is consistent with $\lambda=0$ at the 2$\sigma$ level}
\end{figure}

\citet{gaudiwinn2007} have shown that there is strong degeneracy between $v\sin{I_{*}}$ and $\lambda$ for systems with central transits, i.e. low impact parameters. In this case, placing a Gaussian prior on $v\sin{I_{*}}$ and miss-estimating $\Delta v_{\rm{RM}}$ will lead to a miss-estimation of $\lambda$. This is likely the case with XO-4b with an impact parameter of $b=0.230_{-0.078}^{+0.077}$. To constrain $\lambda$ we apply a Gaussian prior on $v\sin{I_{*}}=8800\pm500$~[m/s] taken from MC08. However, the OTS model requires a higher $v\sin{I_{*}}_{\rm{OTS}}=100_{-420}^{+430}$ and lower $\lambda_{\rm{OTS}}=-20.6_{-8.0}^{+9.0}$ to fit the data. The Hirano model is closer to the stellar prior with $v\sin{I_{*}}_{\rm{Hirano}}=8680_{-440}^{+460}$ and prefers a higher spin-orbit misalignment of $\lambda_{\rm{Hirano}}=-40.0_{-7.5}^{+8.8}$. Figure~\ref{vsinilambda} illustrates the $v\sin{I{*}}$-$\lambda$ degeneracy and where the two models lie in this parameter space. We note that, although the two models agree at nearly $1\sigma$, the implications are very different. In one case, one could infer that the system is almost consistent with being aligned ($\lambda_{\rm{OTS}}$ is consistent with zero at $\sim2\sigma$), whereas this is much less likely for the other model. So, the difference due to the choice of models is not simply quantitative, but is also qualitative. Given the effective temperature of the host star of $T_{\rm{eff}}\sim6400$~K, the inference that XO-4b's orbit is misaligned with its host star would be consistent with the observed trend that hot Jupiters orbiting hot ($T_{\rm{eff}}>6250$~K) stars tend to have high obliquities \citep{winn2010a,albrecht2012}.

Ultimately, we believe the \citet{hirano2010} model and inferences to be more reliable.  This poses a potential complication for future modeling. While the updated \citet{ohta2009} models can be generally applied to RM measurements, the \citet{hirano2010} model requires estimating the model parameters $p$ and $q$, which in turn requires performing the cross-correlation analysis and ultimately having access to the RV data itself or relying on other groups to perform this analysis and publish their results. Without access to the RV data or to the derived values of $p$ and $q$, one must adopt a general model, such as the Ohta model, which may lead to biased and/or incorrect inferences.

\subsubsection{Transit Timing Variations}

We also fit for the mid-transit times of each observed transit, in order to investigate the presence of TTVs. During this we are also able to refine the orbital period by fitting the observed mid-transit times shown in Table~\ref{tabtimes} with a linear function
\begin{eqnarray}\label{eqttv}
T_{c}(E)=T_{c}(0)+EP,
\end{eqnarray}
where E is the Epoch.% The refined transit ephemeris for XO-4b is $P=4.1250723\pm0.0000024$~days with a reduced fit of $\chi^{2}/dof=6.60$.

\begin{deluxetable}{cccccc}[r]
\centering
\tablecaption{Transit Times for XO-4b}
\tablehead{\colhead{Epoch} & \colhead{$T_{C}$} & \colhead{Error} & \colhead{O-C} & \colhead{$\frac{\mathrm{O-C}}{\mathrm{Error}}$}& \colhead{Group\footnotemark[a]}\\
 & (\bjdtdb)& (s) & (s)& &}
\startdata
 -77 & 2454407.535746  & 266 & -1971.90  & -7.40 & X\\
 -76 & 2454411.686963  & 268 &   287.14  &  1.07 & X\\
 -74 & 2454419.936177  & 100 &   207.01  &  2.06 & X\\
 -61 & 2454473.554962  & 140 &  -409.37  & -2.92 & X\\
 -60 & 2454477.685305  & 105 &    46.15  &  0.44 & X\\
 -58 & 2454485.932467  & 175 &  -211.27  & -1.21 & X\\
 -58 & 2454485.938725  & 201 &   329.42  &  1.64 & X\\
 -58 & 2454485.933786  & 176 &   -97.31  & -0.55 & X\\
 -53 & 2454506.560261  &  77 &    -0.41  & -0.01 & X\\
   5 & 2454745.815586  &  59 &   105.39  &  1.77 & F\\
   6 & 2454749.939938  &  63 &    43.29  &  0.69 & F\\
  14 & 2454782.941958  & 102 &   168.95  &  1.65 & D\\
  21 & 2454811.815800  &  97 &    26.14  &  0.27 & U\\
  28 & 2454840.692756  & 104 &   152.38  &  1.45 & U\\
  29 & 2454844.814164  &  95 &  -164.08  & -1.72 & U\\
  30 & 2454848.938339  & 240 &  -241.46  & -1.00 & D\\
  36 & 2454873.688662  & 102 &  -250.21  & -2.44 & D\\
  86 & 2455079.947141  &  91 &   176.96  &  1.94 & F\\
 102 & 2455145.946277  &  89 &     4.57  &  0.05 & D\\
 108 & 2455170.692531  & 129 &  -355.73  & -2.75 & D\\
 124 & 2455236.695540  &  63 &  -193.49  & -3.04 & D\\
 124 & 2455236.698073  &  32 &    25.36  &  0.78 & F\\
 133 & 2455273.825430  & 102 &   174.03  &  1.69 & D\\
 293 & 2455933.834816  &  79 &     7.64  &  0.10 & U\\
\enddata
\footnotetext[a]{X=XOET, F=FLWO, D=DEMONEX, U=UDEM}

\label{tabtimes}
\end{deluxetable}

\begin{figure}[t]
\centering
\includegraphics[width=8cm]{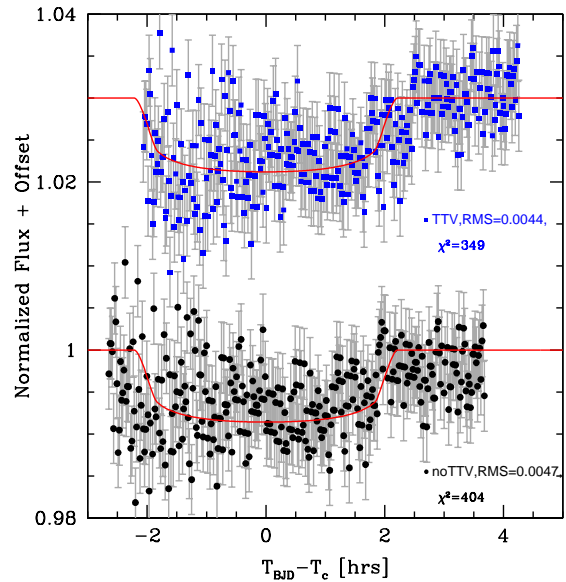}
\caption{\label{20071103} The two light curves for UT 2007-11-03. In blue is the fit where the central transit time is allowed to vary, while the bottom black curve has all of the central transit times fixed to a linear ephemeris. The top curve has a lower residual RMS and $\chi^{2}$, but inferred a $\sim2000$~s transit timing variation.}
\end{figure}

For the night of UT 2007-11-03 (XOET), the best fit mid-transit time results in an O-C of $\sim2000$~s. Upon further investigation, this outlier is likely the result of the noisy photometry from the beginning of the night. UT 2007-11-03 contains no clear ingress, and a feature that could be the egress. After detrending the light curve against time, as no other detrend variables are available, the $\sim2000$~s variation is preferred by EXOFAST when allowing for TTVs. In Figure~\ref{20071103} we show the light curve for the TTV and no TTV cases. When we allow the central transit time to vary, we find that the RMS of the residuals decreases from 0.0047 to 0.0044, and that the $\chi^{2}$ decreases from 404 to 349. 

The data suggest that the $2000$~s TTV solution is the better fit, but a visual inspection of Figure~\ref{20071103} and one could easily choose the no-TTV solution by eye. We do however have a strong prior against $2000$~s transit timing variations and do not believe that the variation is astrophysical in nature, and is perhaps due to a false minimum in the $\chi^{2}$ fit. Little statistical power is contained any single light curve, and to improve the stellar and planetary constraints we omit this night as to not bias the refined transit ephemeris. Omitting this night gives a refined transit ephemeris for XO-4b of
\begin{eqnarray}
P&=&4.1250687\pm0.0000024~\rm{d}\\
T_{C}(0)&=&2454758.18978\pm0.00024~\rm{[\bjdtdb]}
\end{eqnarray} with a reduced fit of $\chi^{2}/dof=3.89$. These values supersede the ephemeris values in Table~\ref{tabmain} from those constrained by the RV data alone during the global fit.

\begin{figure}[t]
\centering
\includegraphics[width=8cm]{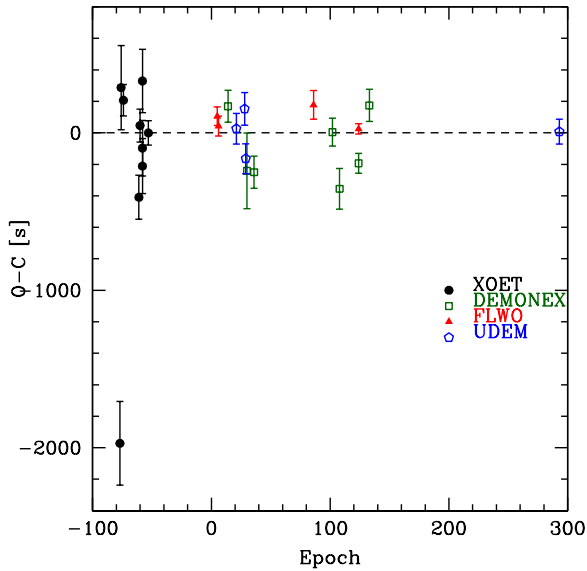}
\caption{\label{ttv0}O-C diagram for the calculated versus measured central transit times, the epochs are taken relative to the RV data. Nights with zero transit timing variations will lie on the dashed line. Even with the exception of UT 2007-11-03 (Epoch -74), there are still a few significant outliers and nights consistent with the presence of transit timing variation. The $\sim2000$~s outlier is the result of poor photometry and the fitting procedure and is omitted in our analysis. The refined transit ephemeris is $T_{0}=2454758.18978\pm0.00024~\rm{[\bjdtdb]}$ and $P=4.1250687\pm0.0000024~\rm{d}$.}
\end{figure}

There are still many nights where the observed mid-transit time is consistent with the presence of TTVs as seen in Figure~\ref{ttv0}. As shown from our detrending analysis it is possible that there are still systematic errors, in addition to our statistical errors, that we have not yet accounted for that could explain the nights that are still consistent with the presence of TTVs. We also note that these are present in all four data sets, not just DEMONEX data.

\section{Results}

In a series of tables we present the median values of the MCMC chains and their 68\% intervals for the stellar parameters, planetary parameters, radial velocity parameters, primary transit parameters, limb-darkening parameters, and secondary eclipse parameters in Table~\ref{tabmain}. The ephemeris in Table~\ref{tabmain} is constrained by the RV data alone, but we do include the refined ephemeris from the mid-transit times as footnotes.

Using the YY2 mass-radius models and the Hirano based RM models we find a refined stellar mass of $1.293_{-0.029}^{+0.030}~\msun$ and stellar radius of $1.554_{-0.030}^{+0.042}~\rsun$ with a companion planet mass of $1.615_{-0.099}^{+0.10}~\mj$ and planet radius of $1.317_{-0.029}^{+0.040}~\rj$. Additionally we find a refined ephemeris of $T_{0}=2454758.18978\pm0.00024$~\bjdtdb and $P=4.1250687\pm0.0000024$~days when fitting for the new ephemeris with the mid-transits as free parameters. We are able to confirm the spin-orbit misalignment of $\lambda=-40.0_{-7.5}^{+8.8}$ when using the Hirano based RM models.

\section{Discussion}

We provide 7 new nights of observations of the misaligned hot Jupiter XO-4b from the DEMONEX telescope and combine that data with previously released data to produce refined stellar and planetary parameters for the XO-4 system analyzed in a homogeneous manner. We investigate a number of possible combinations of detrend parameters and find that the quality of DEMONEX data is significantly improved when detrended against XO-4's position on the CCD and against time. We also note that there is a $3\sigma$ difference in the inferred value of the parameter $R_{P}/R_{*}$ depending on the choice of detrend parameters, and as such, caution should be exercised when detrending data as to not bias results with incorrect inferences.

After investigating both the Torres relation and the Yale-Yonsei isochrones, we are not yet sensitive to the choice of method to resolve the mass-radius degeneracy. We are sensitive to the choice of Rossiter-McLaughlin model used to infer the projected stellar spin-planet orbit angle $\lambda$. We believe the \citet{hirano2010} model and inferences to be more reliable; however, the \citet{ohta2009} models can be generally applied to RM measurements. The \citet{hirano2010} requires having access to the radial velocity data, or a previous estimate of $p$ and $q$ from the cross-correlation analysis. Without such access, one must adopt a general model, such as that based on \citet{ohta2009}, which we have shown may lead to biased and/or incorrect inferences. 

\section{Acknowledgments}

We would like to thank all of the authors who provided their data to us both publicly (McCullough et al., Narita et al.) and privately (P. McCullough, K. Todorov, and P. V. Sada). Thank you to R. Siverd and K. Collins for their work upgrading and maintaining the custom version of EXOFAST used here. Thanks to B. J. Fulton for his comments on coding the Hirano-based RM models. Thanks to K. Collins for her help and guidance with AstroImageJ.

S.V.Jr. is supported by the National Science Foundation Graduate Research Fellowship under Grant No. DGE-1343012. B.S.G is supported by National Science Foundation CAREER Grant AST-1056524.

\begin{deluxetable*}{lcc}[r]
\centering
\tablecaption{Median values and 68\% confidence interval for XO-4b.}
\tablehead{\colhead{~~~Parameter} & \colhead{Units} & \colhead{Value}}
\startdata
\sidehead{Stellar Parameters:}
                               ~~~$M_{*}$\dotfill &Mass (\msun)\dotfill & $1.293_{-0.029}^{+0.030}$\\
                             ~~~$R_{*}$\dotfill &Radius (\rsun)\dotfill & $1.554_{-0.030}^{+0.042}$\\
                         ~~~$L_{*}$\dotfill &Luminosity (\lsun)\dotfill & $3.63_{-0.24}^{+0.28}$\\
                             ~~~$\rho_*$\dotfill &Density (cgs)\dotfill & $0.486_{-0.031}^{+0.023}$\\
                  ~~~$\log{g_*}$\dotfill &Surface gravity (cgs)\dotfill & $4.166_{-0.018}^{+0.013}$\\
                  ~~~$\teff$\dotfill &Effective temperature (K)\dotfill & $6390_{-70}^{+69}$\\
                                 ~~~$\feh$\dotfill &Metallicity\dotfill & $-0.040\pm0.030$\\
             ~~~$v\sin{I_*}$\dotfill &Rotational velocity (m/s)\dotfill & $8680_{-440}^{+460}$\\
           ~~~$\lambda$\dotfill &Spin-orbit alignment (degrees)\dotfill & $-40.0_{-7.5}^{+8.8}$\\
\sidehead{Planetary Parameters:}
                                  ~~~$P$\dotfill &Period (days)\dotfill & $4.12473_{-0.00047}^{+0.00061}$\footnote{Superseded by the transit ephemeris: $P=4.1250707\pm0.0000023$~days}\\
                           ~~~$a$\dotfill &Semi-major axis (AU)\dotfill & $0.05485\pm0.00042$\\
                                 ~~~$M_{P}$\dotfill &Mass (\mj)\dotfill & $1.615_{-0.099}^{+0.10}$\\
                               ~~~$R_{P}$\dotfill &Radius (\rj)\dotfill & $1.317_{-0.029}^{+0.040}$\\
                           ~~~$\rho_{P}$\dotfill &Density (cgs)\dotfill & $0.873_{-0.084}^{+0.081}$\\
                      ~~~$\log{g_{P}}$\dotfill &Surface gravity\dotfill & $3.361_{-0.034}^{+0.032}$\\
               ~~~$T_{eq}$\dotfill &Equilibrium temperature (K)\dotfill & $1641_{-23}^{+25}$\\
                           ~~~$\Theta$\dotfill &Safronov number\dotfill & $0.1036_{-0.0068}^{+0.0070}$\\
                   ~~~$\fave$\dotfill &Incident flux (\fluxcgs)\dotfill & $1.646_{-0.089}^{+0.10}$\\
\sidehead{RV Parameters:}
       ~~~$T_C$\dotfill &Time of inferior conjunction (\bjdtdb)\dotfill & $2454485.993_{-0.11}^{+0.082}$\footnote{Superseded by the transit ephemeris: $T_{C}=2454725.18901\pm0.00025$~\bjdtdb}\\
                        ~~~$K$\dotfill &RV semi-amplitude (m/s)\dotfill & $172\pm10.$\\
                           ~~~$K_R$\dotfill &RM amplitude (m/s)\dotfill & $66.4_{-3.4}^{+3.5}$\\
                    ~~~$M_P\sin{i}$\dotfill &Minimum mass (\mj)\dotfill & $1.615_{-0.099}^{+0.10}$\\
                           ~~~$M_{P}/M_{*}$\dotfill &Mass ratio\dotfill & $0.001192_{-0.000071}^{+0.000073}$\\
                   ~~~$u_{1}$\dotfill &RM linear limb darkening\dotfill & $0.3601_{-0.0069}^{+0.0071}$\\
                ~~~$u_{2}$\dotfill &RM quadratic limb darkening\dotfill & $0.3097\pm0.0026$\\
                                   ~~~$\gamma_{1}$\dotfill &m/s\dotfill & $-4.3_{-8.5}^{+8.4}$\\
                                   ~~~$\gamma_{2}$\dotfill &m/s\dotfill & $0.2_{-2.7}^{+2.8}$\\
                                   ~~~$\gamma_{3}$\dotfill &m/s\dotfill & $-1\pm16$\\
                     ~~~$f(m1,m2)$\dotfill &Mass function (\mj)\dotfill & $0.00000229_{-0.00000038}^{+0.00000044}$\\
\sidehead{Primary Transit Parameters:}
~~~$R_{P}/R_{*}$\dotfill &Radius of the planet in stellar radii\dotfill & $0.08712_{-0.00048}^{+0.00050}$\\
           ~~~$a/R_*$\dotfill &Semi-major axis in stellar radii\dotfill & $7.59_{-0.17}^{+0.12}$\\
                          ~~~$i$\dotfill &Inclination (degrees)\dotfill & $88.26_{-0.63}^{+0.61}$\\
                               ~~~$b$\dotfill &Impact parameter\dotfill & $0.230_{-0.078}^{+0.077}$\\
                             ~~~$\delta$\dotfill &Transit depth\dotfill & $0.007589_{-0.000083}^{+0.000087}$\\
                    ~~~$T_{FWHM}$\dotfill &FWHM duration (days)\dotfill & $0.16880\pm0.00051$\\
              ~~~$\tau$\dotfill &Ingress/egress duration (days)\dotfill & $0.01562_{-0.00053}^{+0.00078}$\\
                     ~~~$T_{14}$\dotfill &Total duration (days)\dotfill & $0.18448_{-0.00079}^{+0.00092}$\\
   ~~~$P_{T}$\dotfill &A priori non-grazing transit probability\dotfill & $0.1203_{-0.0018}^{+0.0027}$\\
              ~~~$P_{T,G}$\dotfill &A priori transit probability\dotfill & $0.1432_{-0.0022}^{+0.0033}$\\
                     ~~~$u_{1B}$\dotfill &Linear Limb-darkening\dotfill & $0.514\pm0.012$\\
                  ~~~$u_{2B}$\dotfill &Quadratic Limb-darkening\dotfill & $0.2544_{-0.0080}^{+0.0079}$\\
                     ~~~$u_{1I}$\dotfill &Linear Limb-darkening\dotfill & $0.2064_{-0.0062}^{+0.0063}$\\
                  ~~~$u_{2I}$\dotfill &Quadratic Limb-darkening\dotfill & $0.3091\pm0.0021$\\
                     ~~~$u_{1R}$\dotfill &Linear Limb-darkening\dotfill & $0.2762_{-0.0066}^{+0.0067}$\\
                  ~~~$u_{2R}$\dotfill &Quadratic Limb-darkening\dotfill & $0.3200\pm0.0022$\\
                ~~~$u_{1Sloanz}$\dotfill &Linear Limb-darkening\dotfill & $0.1744_{-0.0059}^{+0.0060}$\\
             ~~~$u_{2Sloanz}$\dotfill &Quadratic Limb-darkening\dotfill & $0.3020_{-0.0022}^{+0.0021}$\\
                     ~~~$u_{1V}$\dotfill &Linear Limb-darkening\dotfill & $0.3601_{-0.0069}^{+0.0071}$\\
                  ~~~$u_{2V}$\dotfill &Quadratic Limb-darkening\dotfill & $0.3097\pm0.0026$\\
\sidehead{Secondary Eclipse Parameters:}
                  ~~~$T_{S}$\dotfill &Time of eclipse (\bjdtdb)\dotfill & $2454483.931_{-0.11}^{+0.083}$\enddata
\label{tabmain}
\end{deluxetable*}

\end{document}